\documentclass[preprint,aps,floats,prb,fancyhdr,a4paper,superscriptaddress,nofootinbib,showpacs]{revtex4}
\usepackage{latexsym}
\usepackage{amsmath}
\usepackage{amssymb}
\usepackage{amsfonts}
\usepackage{graphicx}
\usepackage{dcolumn}
\usepackage{fancyhdr}
\usepackage[colorlinks]{hyperref}

\begin{document}

\title{A THEORY OF TIME-VARYING CONSTANTS}
\author{Jos\'{e} Antonio Belinch\'{o}n$^{a}$ \& Antonio Alfonso-Faus$^{b}$ \\
{\small $^{a}$Dept. F\'{\i}sica ETS Arquitectura UPM Av. Juan de
Herrera N
4 Madrid 28040 Espa\~{n}a}\\
{\small $^{b}$Dept. Aerotecnia EUIT Aeron\'{a}utica UPM Plaza Cardenal
Cisneros s/n Madrid 28040 Espa\~{n}a}}
\date{September 2000}

\begin{abstract}
We present a flat ($K=0$) cosmological model, described by a perfect fluid
with the ``constants'' $G,c$ and $\Lambda$ varying with cosmological time $t
$. We introduce Planck%
\'{}%
s ``constant'' $\hbar$ in the field equations through the equation of state
for the energy density of radiation. We then determine the behaviour of the
``constants'' by using the zero divergence of the second member of the \emph{%
modified} Einstein%
\'{}%
s field equations i.e. $div(\frac{G}{c^{4}}T_{i}^{j}+\delta_{i}^{j}%
\Lambda)=0,$ together with the equation of state and the Einstein
cosmological equations. Assuming realistic physical and mathematical
conditions we obtain a consistent result with $\hbar c=constant$. In this
way we obtain gauge invariance for the Schr\"{o}dinger equation and the
behaviour of the remaining ``\emph{constants}''.\newline
\end{abstract}
\maketitle

\section{Introduction.}

We study the behavior of the $G,c$ and $\Lambda$ constants in the frame
described by a model with an energy-momentum tensor defined by a perfect
fluid, considering the symmetry implied by a Friedman-Robertson-Walker
metric (we restrict to the flat 3-space case, $K=0$, as has been recently
observed in the CMBR see \cite{B}).

In order to incorporate in the field equations constants like Planck's $%
\hbar $, we consider for the radiation-dominated case an energy density
given by the equation of state
\begin{equation}
\rho=a\theta^{4}
\end{equation}
where $a$ represents the radiation constant and $\theta$ the temperature.
Using this expression for the energy density we can get a possible time
variation for $\hbar$ from the field equations.

We use the modified Einstein's field equations (see \cite{A}, \cite{F} and
\cite{T1}) considering the possibility that all constants may vary with time
$t$, and only with time as inferred from the homogeneity and isotropy of the
metric. We then impose a zero value for the covariant divergence of the
second member of the field equations
\begin{equation}
div\left( T_{i}^{j}+\delta_{i}^{j}\Lambda\right) =0
\end{equation}
This equation, together with the equation of state and the field equations,
will give us the behavior of the ``constants''. To solve these equations we
impose some restrictions and this will drive us to various possibilities.
And we do not consider starting with the condition of zero covariant
divergence for the energy- momentum tensor. This condition has been analyzed
(in a phenomenological way) by various authors (see \cite{H}) that explain
in this way the creation of matter and entropy as due to the time-variation
of $G,c$ and $\Lambda$ only.

Our purpose is not to study thermodynamic properties of the model, but to
determine it including $\hbar$ in the field equations. We impose two
physical and mathematical realistic conditions:

\begin{itemize}
\item  It will always be true the $G$ and $c$ relation
\begin{equation}
\dfrac{G}{c^{2}}=const.
\end{equation}
which is a general covariance principle. We know that this condition is
verified (not imposed) in the case of a bulk viscous fluid. And we restrict
ourselves to the only case such that the parameter controlling the viscosity
describes a behavior topologically equivalent to the case given by the
classical FRW applied to a perfect fluid (our case) (see \cite{T1} and \cite
{T3}).

\item  We impose the following relation for the cosmological ``constant''
obtained through the similarity group that applies to the equations (a
mathematical condition, as a self-similarity relation)
\begin{equation}
\Lambda \propto \dfrac{1}{c^{2}t^{2}}
\end{equation}
Another option that we have, which is equivalent to the above as we will
see, and already analyzed in the literature, is the following one (see \cite
{C}):
\begin{equation}
\Lambda \propto f^{-2}
\end{equation}
\bigskip
\end{itemize}

The paper is organized as follows: In section 2 we present the equations
that define the model and describe the different simplifying hypothesis to
be imposed in the subsequent sections. In section 3 we study a model without
the condition of zero covariant divergence for the energy-momentum tensor.
We will see that with the results obtained the fine structure constant $%
\alpha$ continues being constant (see \cite{V}), despite the time-variation
of all the constants in it. In section 4 we will also see that
Schr\"{o}dinger and Maxwell's equations remain gauge invariant (see \cite{M}%
) with the time varying functions found for the constants in our study. Also
our model does not have the Planck's problem.

\section{\textbf{The Model.}}

\subsection{The equations:}

We will use the field equations in the form:

\begin{equation}
R_{ij}-\dfrac{1}{2}g_{ij}R=\dfrac{8\pi G(t)}{c^{4}(t)}T_{ij}+\Lambda
(t)g_{ij}   \label{ECU1}
\end{equation}

Where the energy momentum tensor is:
\begin{equation}
T_{ij}=\left( \rho+p\right) u_{i}u_{j}-pg_{ij}
\end{equation}
where $p=\omega\rho$ in such a way that $\omega\in\left[ 0,1\right] .$

The cosmological equations are now:
\begin{equation}
2\frac{f\,^{\prime\prime}}{f\,}+\frac{(f\,^{\prime})^{2}}{f\,^{2}}=-\frac{%
8\pi G(t)}{c(t)^{2}}p+c(t)^{2}\Lambda(t)   \label{p1}
\end{equation}
\begin{equation}
3\frac{(f\,^{\prime})^{2}}{f\,^{2}}=\frac{8\pi G(t)}{\,c(t)^{2}}\rho
+c(t)^{2}\Lambda(t)   \label{p2}
\end{equation}
where $f$ stands for the scale factor. Applying the covariance divergence to
the second member of equation (\ref{ECU1}) we get:
\begin{equation}
T_{i;j}^{j}=\left( \dfrac{4c_{,j}}{c}-\dfrac{G_{,j}}{G}\right) T_{i}^{j}-%
\dfrac{c^{4}(t)\delta_{i}^{j}\Lambda_{,j}}{8\pi G}
\end{equation}
that simplifies to:
\begin{equation}
\underset{T_{i;j}^{j}}{\underbrace{\rho^{\prime}+3(\omega+1)\rho H}}=-%
\underbrace{\dfrac{\Lambda^{\prime}c^{4}}{8\pi G}-\rho\dfrac{G^{\prime}}{G}%
+4\rho\dfrac{c^{\prime}}{c}}
\end{equation}
where $H$ stands for the Hubble parameter ($H=\dfrac{f^{\prime}}{f}).$ The
last equation may be written in the form:

\begin{equation}
\rho^{\prime}+3(\omega+1)\rho H+\frac{\Lambda^{\prime}c^{4}}{8\pi G}+\rho
\frac{G^{\prime}}{G}-4\rho\frac{c^{\prime}}{c}=0   \label{def0}
\end{equation}
or the equivalent
\begin{equation}
\dfrac{\rho^{\prime}}{\rho}+3(\omega+1)H+\frac{\Lambda^{\prime}c^{4}}{8\pi
G\rho}+\frac{G^{\prime}}{G}-4\frac{c^{\prime}}{c}=0   \label{def1}
\end{equation}

In this equation we can take a general energy density $\rho$ or use the
black body equation of state $\rho=a\theta^{4}$ where $a=\dfrac{%
\pi^{2}k_{B}^{4}}{15c^{3}\hbar^{3}}$ so that equation (\ref{def1}) is now:
\begin{equation}
4\dfrac{\theta^{\prime}}{\theta}-3\left[ \dfrac{c^{\prime}}{c}+\dfrac
{\hbar^{\prime}}{\hbar}\right] +3(\omega+1)H+\frac{15\Lambda^{\prime}c^{7}%
\hbar^{3}}{8\pi^{3}Gk_{B}^{4}\theta^{4}}+\frac{G^{\prime}}{G}-4\dfrac{%
c^{\prime}}{c}=0   \label{BER4}
\end{equation}

\subsection{Simplifying hypothesis}

\subsubsection{Hypothesis number one (H1).}

We introduce
\begin{equation}
\dfrac{G}{c^{2}}=const.\equiv B
\end{equation}
with $\left[ B\right] =LM^{-1}.$ In this way we obtain directly a term that
controls the behavior of the energy density.
\begin{equation}
\rho=\dfrac{b}{Bt^{2}}\qquad b\in\mathbb{R}
\end{equation}

If we take $\rho=\dfrac{b}{Bt^{2}}$ and introduce $\rho=a\theta^{4}$ we
arrive at
\begin{equation}
\dfrac{b}{Bt^{2}}=a\theta^{4}\Longrightarrow\theta=\left( \dfrac{b}{Ba}%
\right) ^{1/4}t^{-1/2}
\end{equation}

\subsubsection{Hypothesis number two (H2)}

We can make a previous hypothesis on the behavior of the Planck%
\'{}%
s constant as follows:

\begin{itemize}
\item  $c\hbar=const.$ or equivalently $\hbar=\dfrac{A}{c},$ where $A$ is a
proportionality constant. We are motivated to this relation because it
leaves the radiation constant $a$ as constant with a constant $k_{B}=const.$
(see \cite{F})$\medskip$

\item  $\hbar=Fc$ where $\left[ F\right] =L^{1}M^{1}$. This hypothesis is
introduced by some author to explain a possible variation of the fine
structure constant. We will see that such hypothesis is irreconciliable with
our model (see \cite{A}).\bigskip\
\end{itemize}

\subsubsection{Hypothesis number three (H3)}

As a mathematical possibility (self-similar relation), obtained by means of
the similarity group acting on the equations, we will assume the following
behavior for the cosmological ``constant'':
\begin{equation}
\Lambda=\dfrac{d}{c^{2}(t)t^{2}}
\end{equation}
where $d\in\mathbb{R}.$ This hypothesis is a strong one since in some way it
forces to the relation $f(t)=c(t)t$, but this is a desired relation to get
the causality principle preserved.

With this hypothesis we now deal with the equations

\section{Model with $div(T)\neq0$.}

As stated in the introduction this type of model has been recently studied
by different authors. The general idea followed by Harko et al. (see \cite{H}%
) is to study the consequences of the time variation of the constants in
order to solve some problems of the Standard Hot Big-Bang Model (SHBB).

By imposing the condition
\begin{equation}
div(T_{i}^{j})\geq0
\end{equation}
already used by Rastall (see \cite{R}) as of today there has been no proof
that the equality $div(T)=0$ is verified experimentally, hence they impose
the condition $div(T_{i}^{j})\geq a$ where $a$ is a certain positive
function. Rastall imposes a law $div(T_{i}^{j})=\lambda R_{,i\text{ }}i.e.$
proportional to the gradient of the scalar curvature. Hark et al. interpret
the time variation of the constants as responsible for matter creation, and
therefore for entropy, working with the following differential inequality
\begin{equation}
T_{i;j}^{j}=-\dfrac{\Lambda^{\prime}c^{4}}{8\pi G}-\rho\dfrac{G^{\prime}}{G}%
+4\rho\dfrac{c^{\prime}}{c}\geq0
\end{equation}
But in order to solve it they have to impose a previous behavior for the
''constants'' they analyze.

We will follow a different approach here (see \cite{T2}). We will impose
that the covariant divergence of the second member of the Einstein's
equations be zero, i.e.
\begin{equation}
\dfrac{\rho^{\prime}}{\rho}+3(\omega+1)H+\frac{\Lambda^{\prime}c^{4}}{8\pi
G\rho}+\frac{G^{\prime}}{G}-4\frac{c^{\prime}}{c}=0
\end{equation}
and in order to solve it we impose the hypothesis already cited.

\subsection{\textbf{\ }Hypothesis one $\dfrac{G}{c^{2}}=const.\equiv B$
together with hypothesis two\textbf{\ }$\Lambda=\dfrac{d}{c^{2}(t)t^{2}}$%
\textbf{\ }}

In this case the energy density is of a general nature and the equation to
be solved is then (\ref{def1})
\begin{equation}
\dfrac{\rho^{\prime}}{\rho}+3(\omega+1)H+\frac{\Lambda^{\prime}c^{4}}{8\pi
G\rho}+\frac{G^{\prime}}{G}-4\frac{c^{\prime}}{c}=0   \label{BER1}
\end{equation}
which we now analyze.

Since $\dfrac{G}{c^{2}}=const.\equiv B$ then $\rho=\dfrac{b}{Bt^{2}}$ so
that $\dfrac{\rho^{\prime}}{\rho}=\dfrac{-2}{t}$

Taking into account that $\frac{G^{\prime}}{G}=2\frac{c^{\prime}}{c}$ we get
the following expression
\begin{equation}
\frac{\Lambda^{\prime}c^{4}}{8\pi G\rho}=-\dfrac{d}{4\pi b}\left[ \dfrac{%
c^{\prime}}{c}+\dfrac{1}{t}\right]
\end{equation}
Using the field equation
\begin{equation}
3\frac{(f\,^{\prime})^{2}}{f\,^{2}}=\frac{8\pi G}{\,c^{2}}\rho+c^{2}\Lambda
\end{equation}
we obtain $H$ and therefore $f.$%
\begin{equation}
3H^{2}=\dfrac{8\pi Bb}{t^{2}}+\dfrac{d}{t^{2}}\Longrightarrow H=\dfrac
{\varkappa}{t}
\end{equation}
Hence we get $f=K_{\varkappa}t^{\varkappa}$ with $\varkappa=\left( \dfrac{%
8\pi b+d}{3}\right) ^{\dfrac{1}{2}}.$ And rearranging all these results in
the equation (\ref{BER1}) it is obtained:
\begin{equation}
\dfrac{-2}{t}+3(\omega+1)\dfrac{\varkappa}{t}-\dfrac{d}{4\pi b}\dfrac
{c^{\prime}}{c}-\dfrac{d}{4\pi b}\dfrac{1}{t}+2\dfrac{c^{\prime}}{c}-4\dfrac{%
c^{\prime}}{c}=0
\end{equation}
\begin{equation}
\left[ -2-\dfrac{d}{4\pi b}+3(\omega+1)\varkappa\right] \dfrac{1}{t}=\left[
\dfrac{d}{4\pi b}+2\right] \dfrac{c^{\prime}}{c}   \label{REL1}
\end{equation}
simplifying:
\begin{equation}
\dfrac{c^{\prime}}{c}=\left[ \dfrac{12\pi b(\omega+1)\varkappa-8\pi b-d}{%
8\pi b+d}\right] \dfrac{1}{t}
\end{equation}
integrating it we obtain easily:
\begin{equation}
c=K_{\xi}t^{\xi}
\end{equation}
where $\xi=\left[ \dfrac{12\pi b(\omega+1)\varkappa-8\pi b-d}{8\pi b+d}%
\right] .\bigskip$

Another way to look at it is as follows: since we are working with the
constant $B$ when getting $f$ we have obtained another constant $%
K_{\varkappa }$ with dimensions $\left[ K_{\varkappa}\right] =LT^{-\varkappa}
$. With these two dimensional constants we have the solution without the
need of integrating. We can check it with the following result obtained by
means of gauge relations.

The group of governing quantities is $\frak{M=M}\left\{ B,K_{\varkappa
},t\right\} $, and the results are:
\begin{equation}
\begin{array}{l}
c\propto K_{\varkappa}t^{\varkappa-1} \\
G\propto BK_{\varkappa}^{2}t^{2\varkappa-2} \\
\rho\propto B^{-1}t^{-2} \\
f=K_{\varkappa}t^{\varkappa} \\
\Lambda\propto K_{\varkappa}^{-2}t^{-2\varkappa}
\end{array}
\label{OLG2}
\end{equation}
so that we can check the following relations $f=ct,$ $G=Bc^{2}$. We also see
that $\Lambda\propto f^{-2}=c^{-2}t^{-2}$. With the result $\varkappa =%
\dfrac{1}{2}$ we obtain:
\begin{equation}
\begin{array}{l}
c\propto t^{-1/2},\qquad G\propto t^{-1} \\
f\propto t^{1/2},\text{ \ \ \ }\rho\propto t^{-2},\text{ \ \ \ \ }%
\Lambda\propto t^{-1}
\end{array}
\label{OLG1}
\end{equation}
Finally we see that we can regain the law $div(T)=0$ taking into account
that with $f\propto t^{1/2}$ and $\rho\propto t^{-2}$ the relation $\rho
f^{4}=const.$ is satisfied, a relation obtained integrating the equation $%
div(T)=0$. Our results are compatible with this relation.

\subsection{Hypothesis 3 $\Lambda=\dfrac{d}{c^{2}(t)t^{2}}$ together with%
\textbf{\ }$\dfrac{G}{c^{2}}=const.\equiv B$\textbf{\ and }$\protect\rho =a%
\protect\theta^{4}$}

\bigskip Since the constant $B$ has dimensions $\left[ B\right] =LM^{-1}$ we
can get the dimensionless monomia $\pi_{1}=\dfrac{\rho Bt^{2}}{b}$ where $%
b\in\mathbb{R}.$ With these hypothesis and $\pi_{1}$ equation ((\ref{BER4}))
simplifies to:
\begin{equation}
4\dfrac{\theta^{\prime}}{\theta}-3\left[ \dfrac{c^{\prime}}{c}+\dfrac
{\hbar^{\prime}}{\hbar}\right] +3(\omega+1)H-\dfrac{15d}{4\pi^{3}}\frac
{c^{4}\left[ c^{\prime}t+c\right] \hbar^{3}}{Gk_{B}^{4}\theta^{4}t^{3}}+%
\frac{G^{\prime}}{G}-4\dfrac{c^{\prime}}{c}=0   \label{p5}
\end{equation}
that has no immediate integration. We have to take into account the field
equations (\ref{p2}).
\begin{equation}
3H^{2}=8\pi bt^{-2}+dt^{-2}   \label{LUC1}
\end{equation}
from this we get $f=K_{\varkappa}t^{\varkappa}$ where $\varkappa=\left(
\dfrac{8\pi b+d}{3}\right) ^{\dfrac{1}{2}}$ and substituting in (\ref{p5})
together with $G=Bc^{2}$ we get
\begin{equation}
4\dfrac{\theta^{\prime}}{\theta}-3\left[ \dfrac{c^{\prime}}{c}+\dfrac
{\hbar^{\prime}}{\hbar}\right] +\dfrac{3(\omega+1)\varkappa}{t}-\dfrac
{15d}{4\pi^{3}}\frac{c^{2}\left[ c^{\prime}t+c\right] \hbar^{3}}{%
Bk_{B}^{4}\theta^{4}t^{3}}-2\dfrac{c^{\prime}}{c}=0   \label{LUCIL1}
\end{equation}
i.e. one equation with 3 unknowns. Similarly to the previous case we can
take into account the following group of governing quantities $\frak{M=M}%
\left\{ B,K_{\varkappa},t\right\} $ in such a way that
\begin{equation}
c\propto K_{\varkappa}t^{\varkappa-1}
\end{equation}
etc..... (same solution as the previous section, see \ref{OLG2})

If we want to integrate equation (\ref{LUCIL1}) we have to take a decision
on the behavior of he constant $\hbar.$

\begin{enumerate}
\item  Taking $c\hbar=const.\approx\hbar=\dfrac{A}{c}$ then $\dfrac
{\hbar^{\prime}}{\hbar}=-\dfrac{c^{\prime}}{c}$ yielding:
\begin{equation}
4\dfrac{\theta^{\prime}}{\theta}+\dfrac{3(\omega+1)\varkappa}{t}-\dfrac
{15d}{4\pi^{3}}\frac{A^{3}\left[ c^{\prime}t+c\right] }{cBk_{B}^{4}%
\theta^{4}t^{3}}-2\dfrac{c^{\prime}}{c}=0
\end{equation}
Also if $\rho=a\theta^{4}$ and $\rho=\dfrac{b}{Bt^{2}}\Longrightarrow\dfrac
{b}{Bt^{2}}=a\theta^{4}$ we have:
\begin{equation}
k_{B}\theta=\left( \dfrac{15c^{3}(t)\hbar^{3}(t)b}{\pi^{2}Bt^{2}}\right) ^{%
\dfrac{1}{4}}=\left( \dfrac{15A^{3}b}{\pi^{2}B}\right) ^{\dfrac{1}{4}%
}t^{-1/2}
\end{equation}
And substituting into the previous equation we get :
\begin{equation}
\dfrac{-2}{t}+\dfrac{3(\omega+1)\varkappa}{t}-\dfrac{15d}{4\pi^{3}}\frac
{A^{3}\left[ c^{\prime}t+c\right] }{cBt^{3}\dfrac{15A^{3}b}{\pi^{2}Bt^{2}}}-2%
\dfrac{c^{\prime}}{c}=0
\end{equation}
and simplifying
\begin{equation}
\dfrac{-2}{t}+\dfrac{3(\omega+1)\varkappa}{t}-\dfrac{d}{4\pi b}\left[ \dfrac{%
c^{\prime}}{c}+\dfrac{1}{t}\right] -2\dfrac{c^{\prime}}{c}=0   \label{REL2}
\end{equation}
we see that this is the same equations as (\ref{REL1} ) and therefore will
have the same solution \bigskip\ \newline
We can consider another possibility. Take the group of governing quantities $%
\frak{M=M}\left\{ K_{\varkappa },A,t\right\} $ where $K_{\varkappa}$ is the
proportionality constant obtained from $f=K_{\varkappa}t^{\varkappa}$ and $A$
is the constant establishing the relation between $\hbar$ and $c$. The
results obtained by means of the gauge relations are:
\begin{equation}
\begin{array}{l}
G\propto K_{\varkappa}^{6}A^{-1}t^{6\varkappa-4} \\
c\propto K_{\varkappa}t^{\varkappa-1} \\
\hbar\propto K_{\varkappa}^{-1}At^{1-\varkappa} \\
k_{B}\theta\propto K_{\varkappa}^{-1}At^{-\varkappa} \\
\rho\propto K_{\varkappa}^{-4}At^{-4\varkappa} \\
m_{i}\propto K_{\varkappa}^{3}At^{-3\varkappa+2} \\
\Lambda\propto K_{\varkappa}^{-2}t^{-2\varkappa} \\
e^{2}\varepsilon_{0}^{-1}\propto A
\end{array}
\label{OLG3}
\end{equation}

Where $m_{i}$ comes from the energy density definition $\rho_{E}=\dfrac
{nm_{i}c^{2}}{f^{3}}$ ($n$ stands for the particles number) and $%
e^{2}\varepsilon_{0}^{-1}$ from the definition of the fine structure
constant $\alpha$. We can check that we recover the general covariance
property $\dfrac{G}{c^{2}}=const.$ \ iif $\varkappa=\dfrac{1}{2}$. Similarly
we can see that the following relations are satisfied: $\rho=a\theta^{4},$ $%
\rho=Af^{-4}$ (equivalent to $div(T)=0),$ $\Lambda\propto f^{-2\text{ }}$and
$f=ct$ (no horizon problem). And finally $e^{2}\varepsilon_{0}^{-1}\propto
const$. In this way the fine structure constant $\alpha\propto\dfrac{e^{2}}{%
\varepsilon_{0}c\hbar}=const.,$ is a true constant (see \cite{V}). (For a
more detailed analysis on the behavior of the electromagnetic constants see
\cite{T1} and \cite{F1}). With the value $\varkappa=\dfrac{1}{2}$ we get
\begin{equation}
\begin{array}{l}
c\propto t^{-1/2},\text{ \ }\hbar\propto t^{1/2},\text{ \ }G\propto t^{-1},%
\text{ \ }k_{B}\theta\propto t^{-1/2} \\
\text{ \ \ \ \ \ \ \ }f\propto t^{1/2},\text{ \ }\rho\propto t^{-2},\text{ \
\ }m_{i}\propto t^{1/2}
\end{array}
\label{OLG4}
\end{equation}
i.e. same result as in (\ref{OLG1}). We emphasize that all these results
coincide with the work of Midy et al (see \cite{M}). \newline
Another alternative is to consider the governing quantities $\frak{M=M}%
\left\{ B,A,t\right\} $ such that $\left[ A\right] =\left[ A_{\omega}\right]
$ with $\omega=\dfrac{1}{3}$. So that we get the same problem as with the
condition $div(T)=0$ ( Note $\rho=A_{\omega}f^{-3(\omega+1)}$) This would
imply no previous hypothesis on the behavior of $\hbar$.

\item  Assuming $\hbar=Fc$ the equations (\ref{LUCIL1}) is now:
\begin{equation}
4\dfrac{\theta^{\prime}}{\theta}-3\left[ 2\dfrac{c^{\prime}}{c}\right] +%
\dfrac{3(\omega+1)\varkappa}{t}-\dfrac{15d}{4\pi^{3}}\frac{c^{5}\left[
c^{\prime}t+c\right] F^{3}}{Bk_{B}^{4}\theta^{4}t^{3}}-2\dfrac{c^{\prime}}{c}%
=0
\end{equation}
With the relation $\rho=a\theta^{4}$ and $\rho=\dfrac{b}{Bt^{2}}%
\Longrightarrow\dfrac{b}{Bt^{2}}=a\theta^{4}$%
\begin{equation}
k_{B}\theta=\left( \dfrac{15c^{3}(t)\hbar^{3}(t)b}{\pi^{2}Bt^{2}}\right) ^{%
\dfrac{1}{4}}=\left( \dfrac{15c^{6}F^{3}b}{\pi^{2}Bt^{2}}\right) ^{\dfrac{1}{%
4}}
\end{equation}
So that
\begin{equation}
4\dfrac{\theta^{\prime}}{\theta}=6\dfrac{c^{\prime}}{c}-\dfrac{2}{t}
\end{equation}
we obtain:
\begin{equation}
\dfrac{-2}{t}+\dfrac{3(\omega+1)\varkappa}{t}-\dfrac{d}{4\pi b}\left[ \dfrac{%
c^{\prime}}{c}+\dfrac{1}{t}\right] -2\dfrac{c^{\prime}}{c}=0
\end{equation}
Same solution as before (\ref{REL2}). We are getting the same equation, and
the same solution, because we are imposing $\rho\propto t^{-2}$.\bigskip
\newline
If we check this hypothesis with the set of governing quantities $\frak{M=M}%
\left\{ B,F,t\right\} $ we get:
\begin{equation}
\begin{array}{l}
G\propto B^{2}Ft^{-2} \\
c\propto B^{1/2}F^{1/2}t^{-1} \\
f=const. \\
\rho\propto B^{-1}t^{-2} \\
k_{B}\theta\propto B^{1/2}F^{3/2}t^{-2} \\
\hbar\propto B^{1/2}F^{3/2}t^{-1}
\end{array}
\label{OLG5}
\end{equation}
\end{enumerate}

As we see these results are discouraging. In the previous results all the
hypothesis were compatible. Here they are not.\bigskip

\section{Gauge invariance.}

We will check the gauge invariance of the Schr\"{o}dinger, Klein-Gordon and
Maxwell equations (see \cite{PET}).

\begin{enumerate}
\item  Consider Schr\"{o}dinger equation:
\begin{equation}
-\dfrac{h^{2}}{2m}\triangle\psi+U\psi=i\dfrac{h}{2\pi}\partial_{t}\psi
\end{equation}

\item  Klein-Gordon equation:
\begin{equation}
\triangle^{2}\psi+\dfrac{1}{c^{2}}\dfrac{\partial^{2}\psi}{\partial t^{2}}+%
\dfrac{m^{2}c^{2}}{\hbar^{2}}\psi=0
\end{equation}

\item  Maxwell equations are ( see\cite{LL}):
\begin{equation}
rotB=\dfrac{1}{c^{2}}\partial_{t}E\qquad rotE=-\partial_{t}B
\end{equation}
\begin{equation}
divB=0\qquad divE=\dfrac{\rho_{Q}}{\varepsilon_{0}}
\end{equation}
\bigskip(One of us (A,A-F) have deduced a modified Maxwell equations that
remains gauge invariant too, for a more detailed study see \cite{F1})
\end{enumerate}

These equations are all independent from one to another. By introducing a
characteristic length $L$ and time $T$ we can write all these equation in a
dimensionless way.

\begin{itemize}
\item  Schr\"{o}dinger equation is then :
\begin{equation}
l=L\zeta\qquad\nabla=\dfrac{1}{L}\delta\qquad t=T\tau
\end{equation}
\begin{equation}
U=\dfrac{h^{2}}{2mL^{2}}u
\end{equation}
and therefore:
\begin{equation}
-\dfrac{h^{2}}{2mL^{2}}\left( \delta^{2}\psi+u\psi\right) =i\dfrac{h}{2\pi T}%
\partial_{\tau}\psi
\end{equation}

\item  Klein-Gordon equation
\begin{equation}
\frac{\delta^{2}\psi}{L^{2}}+\frac{1}{c^{2}T^{2}}\partial_{\tau}\psi +\frac{%
m^{2}c^{2}}{\hbar^{2}}\psi=0
\end{equation}

\item  While Maxwell%
\'{}%
s equations give
\begin{equation}
B=B^{\ast}\beta\qquad E=E^{\ast}\epsilon\qquad\rho_{Q}=\dfrac{Q}{%
\varepsilon_{0}R^{3}}\varrho_{Q}
\end{equation}
\begin{equation}
\dfrac{B^{\ast}}{L}\delta\times\beta=\dfrac{E^{\ast}}{cT}\dfrac{\partial
\epsilon}{\partial\tau}\qquad\delta\beta=0
\end{equation}
\begin{equation}
\dfrac{E^{\ast}}{L}\delta\times\epsilon=-\dfrac{B^{\ast}}{cT}\dfrac
{\partial\beta}{\partial\tau}\qquad\dfrac{E^{\ast}}{L}\delta\epsilon -4\pi%
\dfrac{Q}{\varepsilon_{0}L^{3}}\varrho_{Q}=0
\end{equation}
\end{itemize}

The invariance of all these equations is ensured if:\bigskip

\begin{enumerate}
\item  $\dfrac{\hbar T}{mL^{2}}=const.$ i.e. $\dfrac{T}{L^{2}}\approx\dfrac
{h}{m}$

\item  $L\thickapprox cT,$ \ \ $mc\thickapprox\dfrac{\hbar}{L},$ \ \ \ $%
mc^{2}\thickapprox\dfrac{\hbar}{T}$

\item  $L\approx cT$ and \ $\dfrac{E^{\ast}}{L}\approx\dfrac{Q}{\varepsilon
_{0}R^{3}}$ \ \ A relation that is consistent with the definition of an
electric field due to an electric charge $Q$.
\end{enumerate}

As we see these are the results obtained elsewhere (\ref{OLG4}) i.e.
\begin{equation}
f\propto t^{1/2}\qquad m\approx h\propto t^{1/2}
\end{equation}
.

We can check now that our model does not have the so called Planck%
\'{}%
s problem. The system behaves now as:
\begin{equation}
\begin{array}{l}
l_{p}=\left( \frac{G\hbar}{c^{3}}\right) ^{1/2}\approx f(t) \\
m_{p}=\left( \frac{c\hbar}{G}\right) ^{1/2}\approx f(t) \\
t_{p}=\left( \frac{G\hbar}{c^{5}}\right) ^{1/2}\approx t
\end{array}
\end{equation}
And since the radius of the Universe $f$ at Planck%
\'{}%
s epoch coincides with the Planck%
\'{}%
s length
\begin{equation}
f(t_{p})\approx l_{p}
\end{equation}
and the energy density at Planck%
\'{}%
s epoch coincides with Planck%
\'{}%
s energy density
\begin{equation}
\rho(t_{p})\approx\rho_{p}\approx t^{-2}
\end{equation}
where $\rho_{p}=m_{p}c^{2}/l_{p}^{3},$ we have no Planck%
\'{}%
s problem here$.$

\section{Conclusions.}

We have presented here a flat cosmological model in which many ``constants''
are considered time-varying. We have found that the only possibility
compatible with the established hypothesis is that the product $c\hbar$ be a
constant. In this way the equations of physics are gauge invariant. The case
considered here is one of a radiation-dominated epoch. However the time
variations found are considered to be of some general value in the sense
that an epoch that is not radiation dominated may conserve some of the
properties found. Further analysis proves that this is the case and that the
result $f=t^{1/2}$ has a general validity.\bigskip

Despite of starting with conditions different from $div(T)=0$ we have found
that this relation is verified for the quantities of our model. With these
results we can say that we can consider adiabatic creation mechanisms of
matter in order to get rid of the entropy problem. In our model the main
quantities present a similar behavior as the FRW classical models, but we
have the advantage of no such problems as the horizon one. The behavior
found for the fine structure constant is that it is a true constant, and
that we can get rid of the Planck's problem. The horizon problem is solved
as soon as the relation $f=ct$ is satisfied. Therefore the Universe expands
at the speed of light.

\end{document}